\documentclass[9pt,twocolumn,twoside]{osajnl}
\journal{ol} % Choose journal (ao, aop, josaa, josab, ol, pr)

\setboolean{shortarticle}{true}
\title{Dual-probe 1D hybrid fs/ps rotational CARS for simultaneous single-shot temperature, pressure, and O$_{2}$/N$_{2}$ measurements}

\author[1,*]{David Escofet-Martin}
\author{Anthony O. Ojo}
\author{Joshua Collins}
\author{Nils Torge Mecker}
\author{Mark Linne}
\author{Brian Peterson}

\affil{Institute for Multiscale Thermofluids, University of Edinburgh, Edinburgh, UK.}
\affil[*]{Corresponding author: david.escofet@ed.ac.uk}

\begin{abstract}
We employ dual-probe one-dimensional (1D) femtosecond (fs)/picosecond (ps) hybrid rotational coherent anti-Stokes Raman spectroscopy (HRCARS) to investigate simultaneous temperature, pressure, and O$_{2}$/N$_{2}$ measurements for gas-phase diagnostics. The dual-probe HRCARS technique allows for simultaneous measurements from the time and frequency-domain. A novel approach for measuring pressure, which offers high accuracy (<1\%) and precision (0.42\%), is presented. The technique is first demonstrated in a chamber for a range of pressures (1-1.5 bar). This technique shows an impressive capability of resolving 1D pressure gradients arising from a N$_{2}$ jet impinging on a surface, both in laminar and turbulent conditions. The technique is shown to be capable of resolving single-shot pressure gradients (0.04 bar/mm) originating from kinetic energy conversion to pressure and resolves characteristic O$_2$/N$_2$ structures from laminar and turbulent mixing.

\end{abstract}

\setboolean{displaycopyright}{true}

\begin{document}

\maketitle

 Temperature is a key thermodynamic variable in reactive flows, because chemical reaction rates are exponentially temperature dependent. Spatially resolved pressure relates to molecular motion and forces, and it is critical for aerodynamic studies. Relative species concentrations are essential when evaluating turbulent mixing, which is prevalent in many practical applications where two gases have to be combined. There is thus a need for spatially resolved, one-dimensional (1D) gas-phase measurements of pressure, temperature and species, which would support flow simulations and optimize operation of system components in practical devices such as gas turbines.

Laser techniques for pressure and temperature are non-intrusive \cite{Chang1990}. In gas-phase thermometry, coherent anti-Stokes Raman spectroscopy (CARS) is esteemed for measurement precision and accuracy \cite{Roy2010}. Historically, nanosecond (ns) CARS provided zero-dimensional (0D) temperature measurements through highly resolved spectra \cite{Beyrau2003}, and simultaneous temperature and pressure measurements were performed by utilizing different spectral features in the frequency-domain \cite{Foglesong1998,Farrow1987}.

New technological developments such as high energy picosecond (ps) and femtosecond (fs) lasers have granted the extension of CARS measurements from 0D to 1D \cite{Kliewer2011} and to two-dimensional (2D) \cite{Bohlin2013a}. These developments permitted time-domain experiments to resolve collisional behavior. Time-resolved CARS has been used to infer Raman coherence decay information \cite{Knopp2002,Morozov2003} and temperature \cite{Roy2008,Seeger2009}.

 Encoding time-domain information into the frequency-domain has been done by chirped-probe-pulse (CPP) CARS \cite{Richardson2011} but remains complex and limited to 0D. Recently, Prince et al \cite{Prince2006} introduced  hybrid fs/ps CARS (HCARS). This technique employs fs beams for the pump/Stokes and a ps beam for the probe. With fs/ps HCARS, frequency-domain information is time-resolved due to the short ps probe, while time-domain information can be obtained by scanning the probe delay \cite{Bohlin2012, Miller2011a}. The significance of this new configuration can be observed in measurements applied in a wide range of pressures, temperatures, and species \cite{Miller2011b,TorgeMecker2020,Miller2015,Bohlin2012,Miller2011a,Kulatilaka2010,Kliewer2012a}. 
Multiple-probe HCARS experiments have been demonstrated in 0D to correct temperature for time-varying collisional environments \cite{Patterson2013}. The  frequency-domain pressure dependence of HCARS has been exploited % 
to measure pressure \cite{Kearney2015a}. Using this pressure dependence and multiple probes, simultaneous pressure and temperature measurements were performed in 0D \cite{Dedic2019a} and 1D \cite{Kearney2020}. However, a trade-off between pressure sensitivity and signal level exists when evaluating pressure \cite{Kearney2020}. This limits the measurements accuracy and precision. Single-shot temporally and spectrally resolved coherences have been demonstrated in 0D \cite{Hosseinnia2020a}, limitations in detector dimensionality complicate 1D measurements.

This Letter introduces a new dual-probe 1D hybrid fs/ps rotational CARS (HRCARS) approach for simultaneous single-shot temperature, pressure, and O$_{2}$/N$_{2}$ ratio measurements. This analysis exploits the strong time-domain pressure dependence of the CARS signal for accurate and precise pressure measurements and overcomes challenges within the frequency-domain. 

The dual-probe 1D fs/ps HRCARS technique is demonstrated here in two environments. The first is an optically accessible chamber \cite{Escofet-Martin2020} to evaluate the accuracy of the pressure measurements. The second is a high speed N$_{2}$ jet impinging on a surface, to demonstrate the utility of the technique in capturing 1D pressure gradients.

\begin{figure}[htbp]
	\centering
	\includegraphics[width=\linewidth]{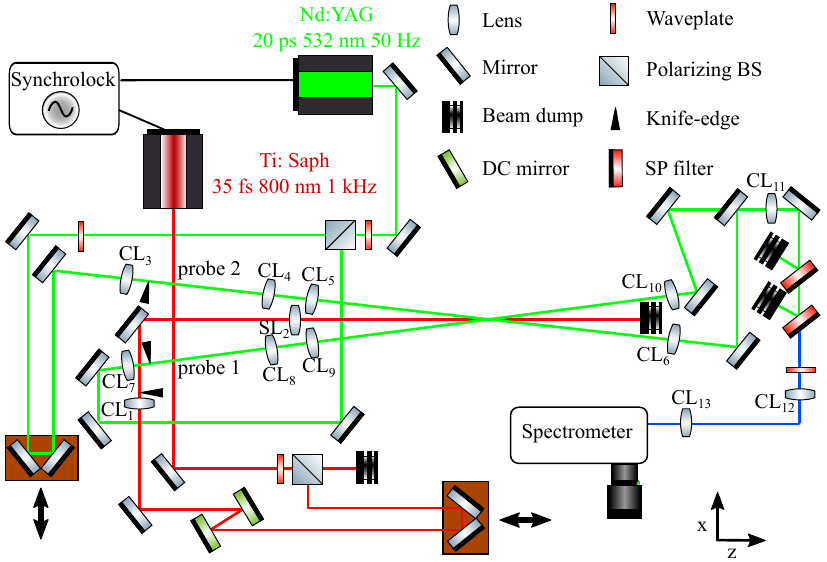}
	\caption{Schematic of the experimental setup for the N$_2$ impinging jet measurements.}
	\label{fig:ssetup}
\end{figure}

Figure \ref{fig:ssetup} illustrates the experimental setup for the N$_2$ impinging jet. The setup is modified from previous HRCARS experimental layouts \cite{Bohlin2013a} to include two probe beams. The seed (Vitara) of a Ti:sapphire amplifier delivering 35 fs pulses with 7.5 mJ at 1 kHz is locked to the ninth harmonic of the oscillator frequency (80 MHz) of a diode pumped Nd:YAG providing 20 ps pulses with 20 mJ at 50 Hz.

\begin{figure}[htbp]
 	\centering
 	\includegraphics[width=\linewidth]{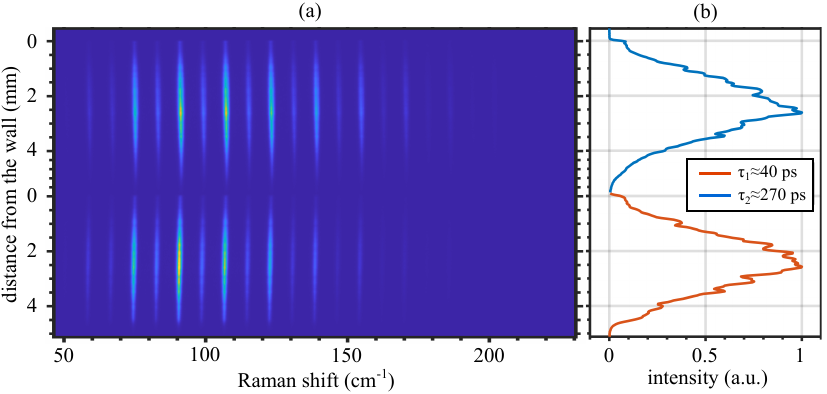}
 	\caption{(a) Single-shot spatially and temporally resolved spectrogram of the dual-probe fs/ps HRCARS near the symmetry axis for the laminar N$_{2}$ impinging jet (374x1024 pixels). (b) Single-shot spatially and temporally resolved projected intensity of the dual-probe fs/ps HRCARS.}
 	\label{fig:singleshot}
 \end{figure}

The pump/Stokes energy is adjusted using a variable attenuator [waveplate plus polarizing beam splitter (BS)]. We use dispersion compensation (DC) mirrors to precompensate for dispersion effects caused by the cylindrical lens (CL)$_{1}$ and spherical lens (SL)$_{2}$. A compensation of $-$220 fs$^2$ provides a near Fourier-transformed pulse after going through the lenses. 
A combination of a CL$_{1}$ (f$_{1}$=300 mm) and a SL$_{2}$ (f$_{2}$=300mm) create a laser sheet that intersects the probe beam sheet. The probe beam is split (with a waveplate plus polarizing BS) into 2 probes: 5\% to probe 1 ($\tau_{1}\approx\ 40 ps$) and 95\% to probe 2 ($\tau_{2}\approx270\ ps$). Two translation stages are used to control the relative delays between the three beams. Two pairs of CLs$_{3,4}$ (f$_{3,4}$=150 mm) and CLs$_{7,8}$ (f$_{7,8}$=150 mm), with the optical power aligned with the y axis and separated by 2f, are placed in the optical path. The CLs$_{4,8}$ image a knife-edge into the probe volume with a magnification slightly $<$ 1. The knife-edge prevents scattering of the beam from the impinging surface. 
CLs$_{5,9}$ (f$_{5,9}$=250 mm) with the optical power along the x axis create two laser sheets. The laser sheets cross at the probe volume with an angle of 5$^\circ$ with the pump/Stokes.
The beams are then collimated with CLs$_{6,10}$ (f$_{6,10}$=400 mm). The mirrors placed between the probe volume and CL$_{11}$ are positioned such that both path lengths are the same. The probe volume (0.055 x 1.2 x 4.5 mm$^3$, $\Delta x$ x $\Delta z$ x $\Delta y$) is relay imaged into the entrance of the spectrometer using CL$_{11}$ and CL$_{12}$ (f$_{11}$=500 mm, f$_{12}$=250 mm). The anti-Stokes and Stokes signals are separated using two angle-tunable short-pass (SP) filters. The CL$_{13}$ (f$_{13}$=100 mm) focuses the light into the spectrometer. The 0.75 m spectrometer is equipped with a 2400 gr/mm grating, which disperses the HRCARS signal onto an electron multiplying charge coupled device (EMCCD). The dispersion is 0.18 cm$^{-1}$/pixel with a spectral resolution of 0.75 cm$^{-1}$ (FWHM). The magnification of the CLs$_{7,8}$ and CL$_{11}$ is set to $\approx$0.8. The EMCCD is binned 2x1 in the y direction and is operated at 25 Hz with 1 ms exposure for a spatial to pixel relationship of 30 $\mu$m/pixel. Figure \ref{fig:singleshot}(a) shows a representative single-shot spectrogram of the CARS intensity, where the two probes are imaged one on top of the other. The resolution of the imaging system is 60 $\mu$m, found by evaluating the sharpness of the imaged knife-edge. A total length of $\approx$4.5 mm is evaluated from the surface. 1D HRCARS spectrograms are corrected for the finite bandwith of the pump/Stokes pulses by scanning the probe beam delay with respect to the pump/Stokes in an argon environment. A weighted average from these delays is used to correct the experimental spectrograms.

For the chamber experiments, the aforementioned knife-edges (Fig. \ref{fig:ssetup}) together with CLs$_{1,3,4,7,8}$ are removed, and SL$_{2}$ is substituted for a CL with f=300 mm. Due to the increased dispersion from the chamber windows (12 mm fused silica), the number of bounces between the DC mirrors is increased to provide -440 fs$^{2}$. The measured pulse after passing through the window is 37 fs (FWHM). The energy employed for the pump/Stokes beam is significantly reduced (<3 mJ) due to supercontinuum light generation in the chamber windows.  

The CARS model used here is detailed in \cite{Yang2017a,Yang2018c}. Pressure dependent linewidths are included with self-broadened linewidths for N$_2$ \cite{Kliewer2012a} and O$_2$-N$_2$ for O$_{2}$ \cite{Millot1992a}. For the frequency-domain fitting, a three-dimensional library with varying temperature, pressure, and O$_2$/N$_2$ ratio was employed. 

The spatially resolved intensities for different probe delays shown in Fig. \ref{fig:singleshot}(b) are key to studying the time-domain pressure fitting. The intensity from integrating the CARS signal is
\begin{equation}
I_{CARS}(T,P,\tau)\propto \int \mathcal{F}\left[P^{(3)}(t,\tau,T,P)\right]d\omega
\label{eq:3}
\end{equation}
where $P^{(3)}$ is the third order polarization
\begin{equation}
P^{(3)}(t,\tau,T,P)\propto \left(\frac{i}{\hbar}\right)^{3}E_{pr}(t-\tau)R_{CARS}(t,T,P)e^{i\omega_{pr}(t-\tau)}.
\label{eq:1}
\end{equation}
$E_{pr}$ is the normalized electric field envelope of the probe, $\tau$ is the relative delay between the probe and the pump/Stokes, and $R_{CARS}$ is the molecular response function
\begin{equation}
R_{CARS}(t,T,P)=\frac{P}{kT}\sum_{J}I_{J;J+2}(T) \times e^{\frac{t}{\hbar}\left(i\Delta E_{J;J+2}-\frac{1}{2}\Gamma_{J;J+2}(T,P)\right)},
\label{eq:2}
\end{equation}
where ${P}/{kT}$ represents the number density effect, $I_{J;J+2}$ is the Raman transition strength, $\Delta E_{J;J+2}$ is the frequency of the transitions, and $\Gamma_{J;J+2}$ is the linewidths.

\begin{figure}[htbp]
	\centering
	\includegraphics[width=\linewidth]{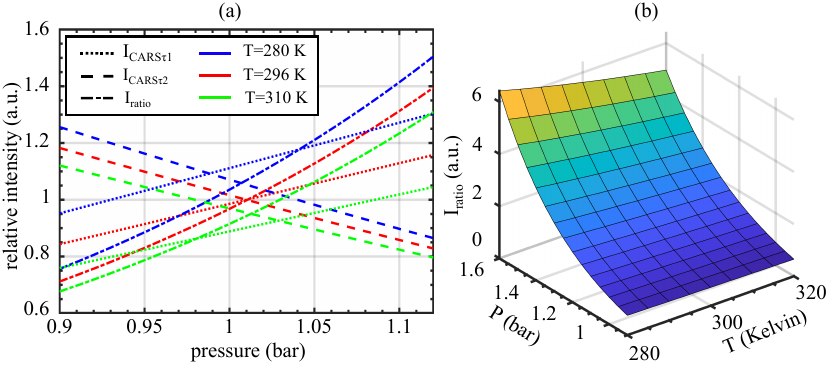}
	\caption{(a) Simulated intensity as a function of pressure (0.9-1.1 bar) and temperature (280-310 K) for early probe ($\tau_{1}= 40$ ps), late probe ($\tau_{2}= 270$ ps) and $I_{ratio}$(b) Simulated relation for I$_{ratio}$, pressure and temperature.}
	\label{fig:ratio}
\end{figure}

Given electric fields for pump/Stokes and probe, the N$_{2}$ HRCARS intensity is a function of probe delay, temperature, and pressure [Eq. (\ref{eq:3})-(\ref{eq:2})]. Figure \ref{fig:ratio}(a) shows the simulated I$_{CARS}$ for two probe delays ($\tau_{1}= 40$ ps, $\tau_{2}= 270$ ps) and the I$_{ratio}$=I${_{CARS\tau1}}$/I${_{CARS\tau2}}$ for a range of pressures of 0.9-1.1 bar and temperatures of 280-310 K. The Raman response ($R_{CARS}(t,T,P)$) contains two pressure dependent terms: number density (${P}/{kT}$) and linewidths ($\Gamma_{J;J+2}(T,P)$). For probe 1 (I${_{CARS\tau1}}$, $\tau_{1}= 40$ ps), the number density contribution dominates the signal intensity for both decreasing temperature and increasing pressure. For probe 2 (I${_{CARS\tau2}}$, $\tau_{2}= 270$ ps), the signal decreases for increasing pressure due to the exponential dependence on linewidth, which overcomes the quadratic signal increase due to number density. 

The I$_{ratio}$ for spatially correlated measurements is not a function of number density or pump/Stokes laser energy, and the pressure sensitivity is increased when compared to any of the individual probes. The relation between the simulated $I_{ratio}$ and the measured $I_{ratio}$ is not complete until a calibration measurement is performed. This accounts for the energy profiles of both individual probe beams, among other things, and is performed separately for chamber and jet experiments, both at ambient pressure (1.01 bar) and temperature. To better utilize the detector dynamic range, the energy split of probes 1 and 2 is adjusted so that $I_{ratio}\approx 1$ under the calibration conditions.   

Figure \ref{fig:ratio}(b) shows the simulated normalized I$_{ratio}$ for a range of temperatures of 280-310 K and pressures of 0.9-1.6 bar. The surface plot illustrates the strong sensitivity of the I$_{ratio}$ with respect to pressure; a variation from 1 to 1.6 bar produces an I$_{ratio}$ change greater than five for this temperature range. The I$_{ratio}$ sensitivity with respect to pressure is predominantly a function of $\Delta \tau=\tau_{2}-\tau_{1}$ and can be adjusted to match the expected measured pressure to the dynamic range of the detector.

The fitting procedure starts by assuming ambient pressure in a frequency-domain fit for probe 1, so we obtain temperature ($T_{1}$). This temperature is then used to obtain pressure using the I$_{ratio}$-pressure relation. The calculated pressure is then used to refit probe 1, resulting in the final temperature (${T}$). With the final temperature and the I$_{ratio}$-pressure relation, we can obtain the final pressure (${P}$) that has already converged. The statistics displayed in Table \ref{tab:calibration} are for spatially and time-resolved measurements in the chamber, from evaluating the central 20 spectral rows in 50 shots. The measurements precision for the chamber is hindered compared to the N$_{2}$ impinging jet experiments due to the reduced pump/Stokes energy ($<$3 mJ).

\begin{table}[htbp]
	\centering
	\caption{\bf Pressure and temperature measurements in the chamber}
	\begin{tabular}{cccccc}
		\hline
		  $\overline{T_{1}}$(K) & $\overline{T}$(K) & $\overline{P}$(bar) &$\frac{\sigma_{T}}{\overline{T}}$ (\%) & $\frac{\sigma_{P}}{\overline{P}}$ (\%) & $\overline{P_{trans.}}$(bar)\\
		\hline
		296.43 & 296.43 & 1.01 & 1.36 & 0.83 & 1.01 \\
		298.59 & 296.59 & 1.22 & 1.14 & 1.29 & 1.21\\
		299.73 & 296.87 & 1.31 & 1.08 & 1.72 & 1.30\\
		300.17 & 296.64 & 1.38 & 1.04 & 1.91 & 1.37\\
		301.30 & 296.83 & 1.48 & 0.94 & 2.44 & 1.49\\
		\hline
	\end{tabular}
	\label{tab:calibration}
\end{table}

 $T_{1}$ shows a systematic increase of 1.7\% as pressure increases 0.47 bar, while the final calculated temperature ($\overline{T}$) only varies 0.2\% in the studied pressure range. The pressure measurements are benchmarked against a pressure transducer ($\overline{P_{trans.}}$). The agreement between $\overline{P}$ with $\overline{P_{trans.}}$ is better than $1$ \%. The temperature precision improves from 1.36 to 0.94\% as pressure increases due to the increased signal from I$_{CARS\tau1}$. The pressure precision decreases from 0.83 to 2.44\% as pressure increases, due to the signal decrease from I$_{CARS\tau2}$.

To demonstrate the utility of the dual-probe HRCARS technique, measurements were applied to a N$_{2}$ jet impinging on a surface. The surface is curved along the z direction (d=300 mm) for easier optical access. Measurements are performed under two conditions, $\overline{V}$=5.2 m/s (Re=1,420, laminar) and $\overline{V}$=104.2 m/s (Re=28,510, turbulent). The jet inner diameter is D=4 mm, and the distance between the jet and the surface is H=11 mm, providing a H/D ratio of 2.75. The 4.5 mm closest to the plate are evaluated for temperature, pressure, and O$_2$/N$_{2}$. For every 2D field, twenty 1D measurements are performed, spaced every 200 $\mu$m, covering the radial direction $\pm$2 mm of the jet impingement. The surface is (4 x 4.5 mm$^2$, $\Delta x$ x $\Delta y$), and each averaged field contains the analysis of (150 x 20 x 100) 0.3 million spectra.

\begin{figure}[htbp]
	\centering
	\includegraphics[width=\linewidth]{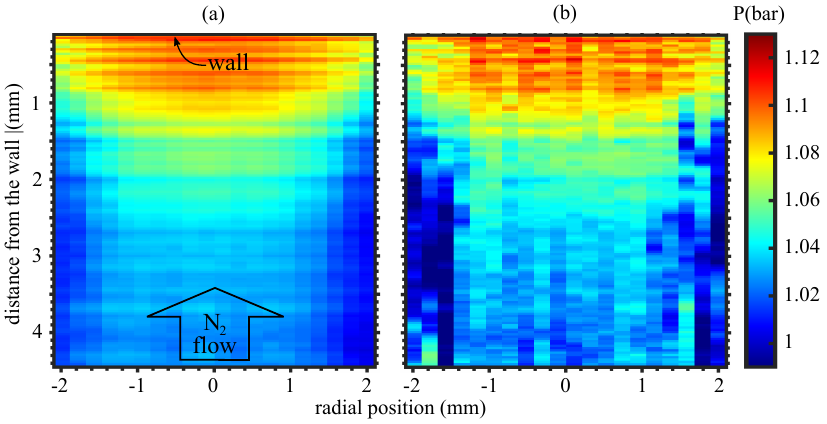}
	\caption{ 2D pressure field from a turbulent (Re=28,510) N$_{2}$ jet impinging on a surface: (a) averaged and (b) single-shot.}
	\label{fig:pressure}
\end{figure}

The laminar pressure field and the laminar and turbulent temperature fields are uniform, thus not shown here. The  pressure and temperature fields obtained from the laminar jet are used to evaluate the measurements precision. The spatio-temporal precision for pressure is 0.42\% ($\sigma_{P}/\overline{P}$) and for temperature is 0.62\% ($\sigma_{T}/\overline{T}$). Figure \ref{fig:pressure} shows averaged and instantaneous turbulent pressure fields. The maximum measured pressure is $\approx$1.11 bar in both the averaged and single-shot fields. This maximum pressure is located near the symmetry axis close to the surface and decays both radially and axially. Spatial pressure gradients of the order of 0.04 bar/mm are well-captured.

\begin{figure}[htbp]
	\centering
	\includegraphics[width=\linewidth]{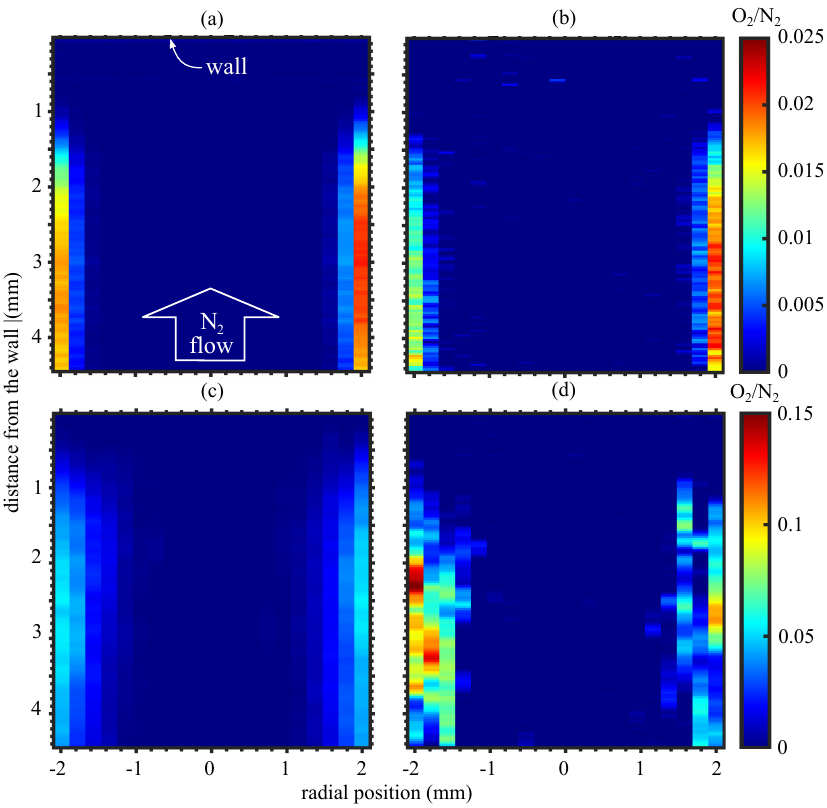}
	\caption{2D O$_{2}$/N$_{2}$ field from a N$_{2}$ jet impinging on a surface: laminar jet [(a) averaged field, (b) single-shot field] and turbulent jet [(c) averaged field, (d) single-shot field].}
	\label{fig:oxigen}
\end{figure}

 To further illustrate the jet behavior, Fig. \ref{fig:oxigen} shows both averaged and single-shot O$_{2}$/N$_{2}$ ratios for both (a), (b) the laminar and (c), (d) the turbulent conditions. The stagnating flow field causes no O$_{2}$ detection close to the plate. Comparing both O$_{2}$/N$_{2}$ averaged fields, the turbulent field has a  O$_{2}$/N$_{2}\approx$ 0.06 ratio, a three-fold increase due to enhanced air entrainment near the jet radius. Differences in mixing mechanisms are captured in the single-shot measurements: for the laminar jet, the O$_{2}$/N$_{2}$ measurements show smooth O$_{2}$/N$_{2}$ gradients typical of mixing dominated mostly by diffusion, while the turbulent jet shows fine structure of air pockets from air entrainment, which also demonstrates the exceptional spatial resolution.

In summary, with dual-probe 1D HRCARS, by extracting information from both the time and frequency-domains, we have demonstrated simultaneous pressure, temperature, and O$_{2}$/N$_{2}$ ratio measurements. A new methodology to evaluate pressure leads to high accuracy (<1\%) for the pressure range 1-1.5 bar, remarkable spatio-temporal temperature precision of 0.62\% ($\sigma_{T}/\overline{T}$), unmatched spatio-temporal pressure precision of 0.42\% ($\sigma_{P}/\overline{P}$), and fine spatial resolution (0.055 x 1.2 x 0.060 mm$^{3}$). The pressure measurements are able to describe pressure gradients of the order of 0.04 bar/mm. The O$_{2}$/N$_{2}$ measurements also capture the spatial characteristics from the mixing interface both in laminar and turbulent conditions. Applying this measurement technique in a more challenging environment with higher pressures, temperatures, and species gradients will be the focus of future work. For such studies, anticipated issues such as reduced signal due to fast collisional dephasing at high pressures, unknown collisional partners, and choice of optimal probe delays will be investigated.

\medskip

\noindent\textbf{Funding.} Engineering and Physical Sciences Research Council (EP/P020593/1, EP/P001661/1); European Research Council (759546).

\medskip

\noindent\textbf{Disclosures.} The authors declare no conflicts of interest.

%\section{References}

%\bigskip
%\noindent Add citations manually or use BibTeX. See \cite{Zhang:14,OSA,FORSTER2007,testthesis,manga_rao_single_2007}.

% Bibliography
\bibliography{library/opticsletters}

% Full bibliography added automatically for Optics Letters submissions; the following line will simply be ignored if submitting to other journals.
% Note that this extra page will not count against page length
\bibliographyfullrefs{library/opticsletters}

%Manual citation list
%\begin{thebibliography}{1}
%\bibitem{Zhang:14}
%Y.~Zhang, S.~Qiao, L.~Sun, Q.~W. Shi, W.~Huang, %L.~Li, and Z.~Yang,
 % \enquote{Photoinduced active terahertz metamaterials with nanostructured
  %vanadium dioxide film deposited by sol-gel method,} Opt. Express \textbf{22},
  %11070--11078 (2014).
%\end{thebibliography}

% Please include bios and photos of all authors for aop articles
\ifthenelse{\equal{\journalref}{aop}}{%
\section*{Author Biographies}
\begingroup
\setlength\intextsep{0pt}
\begin{minipage}[t][6.3cm][t]{1.0\textwidth} % Adjust height [6.3cm] as required for separation of bio photos.
  \begin{wrapfigure}{L}{0.25\textwidth}
    \includegraphics[width=0.25\textwidth]{john_smith.eps}
  \end{wrapfigure}
  \noindent
  {\bfseries John Smith} received his BSc (Mathematics) in 2000 from The University of Maryland. His research interests include lasers and optics.
\end{minipage}
\begin{minipage}{1.0\textwidth}
  \begin{wrapfigure}{L}{0.25\textwidth}
    \includegraphics[width=0.25\textwidth]{alice_smith.eps}
  \end{wrapfigure}
  \noindent
  {\bfseries Alice Smith} also received her BSc (Mathematics) in 2000 from The University of Maryland. Her research interests also include lasers and optics.
\end{minipage}
\endgroup
}

\end{document}